\documentclass[twocolumn,showpacs,preprintnumbers,amsmath]{revtex4}
\usepackage{graphicx}
\usepackage{epsfig}
\usepackage{amsmath,amssymb}

\begin{document}

\title{Extended Pairing Model for Heavy Nuclei}
\author{V. G. Gueorguiev$^{1,2}$, Feng Pan$^{3}$, and J. P. Draayer$^{1}$}
\affiliation
{$^{1}$Department of Physics \& Astronomy, Louisiana State University,
Baton Rouge, LA 70803, USA\\
$^{2}$Institute of Nuclear Research \& Nuclear Energy,
Bulgarian Academy of Sciences, Sofia 1784\\
$^{3}$Department of Physics, Liaoning Normal University,
Dalian, 116029, P. R. China}

\begin{abstract}
We study binding energies in three isotopic chains ($^{100-130}$Sn,
$^{152-181}$Yb, and $^{181-202 }$Pb) using the extended pairing model
with Nilsson single-particle energies. The exactly solvable nature of
the model means that the pairing strength $G(A)$ required to reproduce the
experimental binding energies can be determined uniquely. The valence
space consists of the neutron single-particle levels between two closed
shells corresponding to the magic numbers 50-82 and 82-126. In all three
isotopic chains, $\log(G(A))$ has a smooth quadratic behavior for even as
well as odd nucleon numbers A; $\log(G(A))$ for even and odd A are very
similar. Remarkably, $G(A)$ for all the Pb isotopes can be also described by a two parameter expression that is inversely proportional to the dimensionality of the
model space.
\end{abstract}

\maketitle

\vspace{0.5cm}



In a previous paper we introduced the extended pairing model,
explored its properties, and applied it to some well-deformed nuclei
\cite{FirstPaper}. Metallic clusters of nano-scale size may provide 
another physical system where one can test and study the 
applicability of the extended pairing interaction \cite{metallic 
grains}. In this paper we exploit the exactly solvable nature
of the model to explore its applicability in reproducing experimental
nuclear binding energy of three distinct isotopic chains: $^{100-130}$Sn,
$^{152-181}$Yb, and $^{181-202}$Pb. The results suggest that the model is
useful for its simplicity, its exact solvability, and its ability to track
and predict experimental binding energies of long sequences of nuclei.

For each isotopic chain we take the binding energy of a closed neutron
shell to be our zero-energy reference point. Each such closed neutron
shell nucleus ($^{100}$Sn, $^{152}$Yb, and  $^{208}$Pb) and its odd-$A$
neighbor ($^{101}$Sn, $^{153}$Yb, and $^{207}$Pb) are assumed to be well
described by the independent particle model with Nilsson single-particle
energies; thus no extended pairing interaction terms are needed for these
nuclei. The energy scale of the Nilsson single-particle energies is set so
that the binding energies of $^{101}$Sn, $^{153}$Yb, and $^{207}$Pb are
reproduced by the independent particle model. For all the other nuclei we
solve for the pairing strength $G(A)$ that reproduces the experimental
binding energies exactly within the chosen model space. The valence model
space consists of the neutron single-particle levels between two closed
shells corresponding to the magic numbers 50-82 and 82-126. The structure
of the model space is reflected in the values of $G(A)$. In particular,
in all the cases studied $\log(G(A))$ has a smooth quadratic behavior for
even and odd $A$ with a minimum in the middle of the model space where the
size of the space is a maximal; $\log(G(A))$ for even $A$ and odd
$A$ are very similar which suggests that more detailed
shell-model analyses may result in the same $G(A)$ functional form for even
$A$ and odd $A$ isotopes. In particular, for the Pb isotopes the even and
odd $A$ nuclei are described by a single two-parameter expression for the
pairing strength $G(A)$ that is inversely proportional to the
dimensionality of the valence model space.

To set the stage and establish the notation we begin with a brief review
of the underpinnings of the extended pairing model. The standard Nilsson
plus pairing Hamiltonian is given by
\begin{equation}
\hat{H}=\sum_{j=1}^{p}\epsilon _{j}n_{j}-G\sum_{i,j=1}^{p}a_{i}^{+}a_{j},
\label{eqno1}
\end{equation}
where $p$ is the total number of single-particle levels considered,
$\epsilon _{j}$ are single-particle energies, $G$ is the overall pairing
strength ($G>0$ ), $n_{j}=c_{j\uparrow }^{\dagger}c_{j\uparrow} +
c_{j\downarrow }^{\dagger }c_{j\downarrow}$ is the number operator for
the $j$-th single-particle level, $a_{i}^{+}=c_{i\uparrow }^{\dagger}
c_{i\downarrow}^{\dagger }$ ($a_{i}=(a_{i}^{+})^{\dagger}
=c_{i\downarrow} c_{i\uparrow}$) are pair creation (annihilation)
operators where $c_{j\uparrow}^{\dagger}$ ($c_{j\downarrow}^{\dagger}$)
creates a fermion in the $j$-th single-particle level. The up and
down arrows refer to time-reversed states. Since each Nilsson level can
only be occupied by one pair due to the Pauli Exclusion Principle, the
operators $a_{i}^{+}$, $a_{i}$, and $n_{i}$ form a hard-core boson
algebra:
\begin{equation}
\lbrack a_{i},a_{j}^{+}]=\delta
_{ij}(1-n_{i}),~~[a_{i}^{+},a_{j}^{+}]=0=(a_{i}^{+})^{2}.  \label{eqno2}
\end{equation}

As a generalization of (\ref{eqno1}), we consider the following extended
pairing Hamiltonian \cite{FirstPaper}:
\begin{widetext}
\begin{eqnarray}
\label{eqno3}
\hat{H} &=&\sum_{j=1}^{p}\epsilon _{j}n_{j}-
G\sum_{i,j=1}^{p}a_{i}^{+}a_{j} -G\left( \sum_{\mu =2}^{p
}{\frac{1}{{{(\mu !)}^{2}}}}
\sum_{i_{1}\neq i_{2}\neq \cdots \neq i_{2\mu
}}a_{i_{1}}^{+}a_{i_{2}}^{+}\cdots a_{i_{\mu }}^{+}a_{i_{\mu +1}}a_{i_{\mu
+2}}\cdots a_{i_{2\mu }}\right) .
\end{eqnarray}
\end{widetext}Besides the usual single-particle terms and the standard
pairing interaction (\ref{eqno1}), this interaction includes many-pair
interaction terms which connect configurations that differ by more than one
pair.

The main advantage of the extended pairing model is that it is exactly
solvable  \cite{FirstPaper}. It is easy to see that any term of the
form $a_{i}^{+}\cdots a_{j}^{+}$ that forms eigenstates of (\ref{eqno3})
should enter with different indices
$i\neq \cdots \neq j$. Let $|j_{1},\cdots ,j_{m}\rangle $ denotes a
pairing vacuum
state that satisfies $a_{i}|j_{1},\cdots ,j_{m}\rangle =0$ for $1\leq
i\leq p$ and $i\neq j_{s}$, where $j_{1},j_{2},\cdots ,j_{m}$
indicate those $m$ levels that are
occupied by single nucleons. Any state that is occupied by a single nucleon is blocked to the hard-core bosons due to the Pauli exclusion principle. This means that the
space of all possible configurations decomposes in a direct sum of
orthogonal sub-spaces that are invariant under the action of the
Hamiltonian and are labeled by the positions of the unpaired nucleons.
Thus a $k$-pair eigenstates of (\ref{eqno3}) has the form
\begin{equation}
|k;\zeta ;j_{1},\cdots ,j_{m}\rangle =\sum_{1\leq i_{1}<\cdots <i_{k}\leq
p}C_{i_{1}\cdots i_{k}}^{(\zeta )}a_{i_{1}}^{+}\cdots
a_{i_{k}}^{+}|j_{1},\cdots ,j_{m}\rangle ,  \label{eqno4}
\end{equation}
where $C_{i_{1}i_{2}\cdots i_{k}}^{(\zeta )}$ are expansion coefficients
that need to be determined, and the strict ordering to the indices
$i_{1},i_{2},\cdots ,i_{k}$ reminds us that double occupation is not
allowed. It is always assumed that the level indices $j_{1},~j_{2},\cdots
,j_{m}$ should be excluded from the summation in (\ref{eqno4}). Since the
general formalism is similar, we will focus on the seniority zero case
(no unpaired particles).

In a manner that is similar to the results given by the Bethe ansatz, the
expansion coefficient  $C_{i_{1}i_{2}\cdots i_{k}}^{(\zeta )}$ in
(\ref{eqno4}) can be written as \cite{Bethe ansatz}:
\begin{equation}
C_{i_{1}i_{2}\cdots i_{k}}^{(\zeta )}=\frac{1}{{1-y^{(\zeta )}}
E_{i_{1}...i_{k}}},\quad E_{i_{1}...i_{k}}=\sum_{\mu =1}^{k}2{\epsilon }
_{i_{\mu }}  \label{eqno5}
\end{equation}
where $y^{(\zeta )}$ is a number that is to be determined. To solve (\ref
{eqno3}) using (\ref{eqno4}) and (\ref{eqno5}) one directly applies
Hamiltonian (\ref{eqno3}) on the $k$-pair state (\ref{eqno4}). Using the
hard-core boson algebraic relations given by (\ref{eqno2}), one can
determine the action of the mean-field part of the Hamiltonian
(\ref{eqno3}):
\begin{eqnarray}
\sum_{j}\epsilon _{j}n_{j} &|&k;\zeta ;0\rangle =\frac{1}{{y^{(\zeta )}}}
\times  \label{eqno6} \\
&\times &\left( |k;\zeta ;0\rangle -\sum_{1\leq i_{1}<\cdots <i_{k}\leq
p}a_{i_{1}}^{+}\cdots a_{i_{k}}^{+}|0\rangle \right) ,  \nonumber
\end{eqnarray}
and for the extended pairing part of the Hamiltonian (\ref{eqno3}):
\begin{widetext}
\begin{eqnarray}
\label{eqno7} &&\left( \sum_{i}a_{i}^{+}a_{i}+\sum_{\mu =1}^{\infty
}{\frac{1}{{{(\mu !)}^{2} }}}\sum_{i_{1}\neq i_{2}\neq \cdots \neq
i_{2\mu}}a_{i_{1}}^{+}a_{i_{2}}^{+}\cdots a_{i_{\mu }}^{+}a_{i_{\mu
+1}}a_{i_{\mu+2}}\cdots a_{i_{2\mu }}\right) |k;\zeta ;0\rangle = \\
&=&\left( \sum_{1\leq i_{1}<i_{2}<\cdots <i_{k}\leq p}C_{i_{1}i_{2}\cdots
i_{k}}^{(\zeta )}\right) \sum_{1\leq i_{1}<i_{2}<\cdots <i_{k}\leq
p}a_{i_{1}}^{+}a_{i_{2}}^{+}\cdots a_{i_{k}}^{+}|0\rangle +(k-1)|k;\zeta
;0\rangle . \nonumber
\end{eqnarray}
\end{widetext}By combining (\ref{eqno6}) and (\ref{eqno7}), the
$k$-pair excitation energies of (\ref{eqno3}) are given by
\begin{equation}
E_{k}^{(\zeta )}=\frac{1}{y{^{(\zeta )}}}-G(k-1),  \label{eqno8}
\end{equation}
and the undetermined variable $y^{(\zeta )}$ is given by
\begin{equation}
{\frac{1}{y{^{(\zeta )}}}}+\sum_{1\leq i_{1}<i_{2}<\cdots <i_{k}\leq p}{
\frac{G}{(1-y{^{(\zeta )}E_{i_{1}...i_{k}})}}=0}.  \label{eqno9}
\end{equation}
The additional quantum number $\zeta $ can now be understood as the
$\zeta$-th solution of (\ref{eqno9}). Similar results for many
broken-pair systems can be derived by using this approach except that the
indexes $j_{s}$ of the level occupied by the single nucleons should be
excluded from the summation in (\ref{eqno4}) and the single-particle
energy term $\epsilon _{j_{s}}$ from the first part of (\ref{eqno3})
should be added to the total eigenenergy.

By comparing (\ref{eqno8}) and (\ref{eqno9}) to the exact solutions
of the Heisenberg algebraic Hamiltonian with a one-body interaction
\cite{Feng'99},  one can regard the operator product
$a_{i_{1}}^{+}a_{i_{2}}^{+}
\cdots a_{i_{k}}^{+}$ in (\ref{eqno4}) as a `grand' boson. The
corresponding `single-particle energy' of the `grand' boson is
$E_{i_{1}i_{2}...i_{k}} = \sum_{\mu =1}^{k}2{\epsilon }_{i_{\mu }}$. The
eigenstates (\ref{eqno4}) are not normalized, but can be normalized
once the coefficients $C_{i_{1}i_{2}\cdots i_{k}}^{(\zeta )}$ are known.
The eigenstates (\ref {eqno4}) with different roots given by
(\ref{eqno9}) are mutually orthogonal since they correspond to
eigenstates with different eigenvalues.

The $k$ coupled non-linear equations of the standard pairing model 
\cite{Richardson} are difficult to solve numerically, especially when 
the number of pairs $k$
and number of levels $p$ are large. Specifically, there should
be $\binom{p}{k}=\frac{p!}{(p-k)!k!}$ distinct roots, which can be a
very large number for an entire deformed major shell. While there have
been major advances in methods for solving the associated Richardson
equations, for example \cite{Stephan}, the theory is limited because of
the coupled non-linear nature of the equations. In contrast to this, for
the extended pairing model there is but one variable $y^{(\zeta )}$ and
one equation (\ref{eqno9}); so a relatively simple Mathematica code, for
example, can be used to solve for the roots.

Our Nilsson plus extended pairing model uses single-particle energies of
each nucleus as calculated within the Nilsson deformed shell model with
experimentally evaluated deformation parameters \cite{Nix JR}.
Experimental binding energies are taken from reference \cite{Audi G}.
The theoretical binding energies are calculated relative to a particular
core. We use $^{152}$Yb, $^{100}$Sn, and $^{208}$Pb  as cores in our
calculations. While there are changes in the binding energy of the core
since the corresponding Nilsson levels change as a function of the
deformation, our results indicate that such core affects are an order of
magnitude smaller than the overall even-odd staggering in the binding 
energy. From the binding energy of a nucleus next to the core, we 
calculate an overall energy scale for the
Nilsson single-particle energies. For an even number of neutrons, we
consider only pairs of particles (hard bosons). For an odd number of
neutrons, we apply Pauli blocking of the Fermi level of the last unpaired
fermion and consider the remaining fermions as if they were an even
$A$ fermion system. By using (\ref{eqno8}) and (\ref{eqno9}), values of
$G$ are calculated so that the experimental and theoretical binding energy
match exactly. With the help of (\ref{eqno8}) and (\ref{eqno9}) and some
algebra, one can see that for a given single particle energies there is an upper limit to the value of the binding energy for which a physically meaningful exact solution can be
constructed. This upper value of the binding energy for each nucleus is
given by the energy of the lowest ``grand boson'', namely,
$E_{gb}=\sum_{\mu =1}^{k}2 {\epsilon }_{\mu}$.

\begin{figure}[htbp]
\includegraphics[width=8cm]{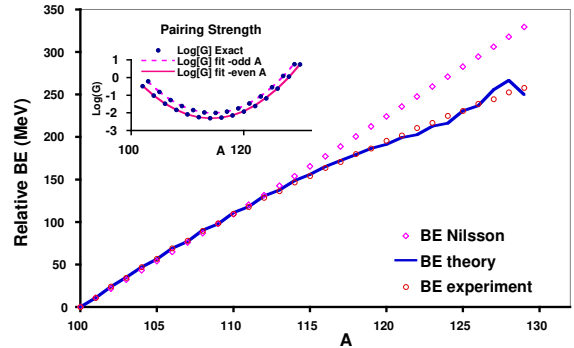}
\caption{The solid line gives the binding energies of the Sn
isotopes relative to that of the $^{100}$Sn core.  The single-particle
energy scale is set by the binding energy of $^{101}$Sn.  The inset shows
the fit to values of $G$ that reproduces the experimental data exactly.
The two fitting functions are: $\log(G(A))=365.0584 - 6.4836 A + 0.0284
A^2$ for even values of $A$ and
$\log(G(A))=398.2277 - 7.0349 A + 0.0307 A^2$ for odd values of $A$.
The Nilsson BE energy is the lowest energy of the non-interacting system.}
\label{Sn-isotopes}
\end{figure}

We now turn to our results for the binding energies and $\log(G)$ values
within the Nilsson plus extended pairing model for three isotopic chains:
$^{100-130}$Sn, $^{152-181}$Yb, and $^{181-202 }$Pb. Figure
\ref{Sn-isotopes} shows the results for the $^{100-130}$Sn isotopes.
Calculations for the pairing strength G were carried out for the
$^{102-130}$Sn isotopes within the 50-82 neutron shell. Once this was
done, a quadratic polynomial fit of the $\log(G)$ values for even and odd
$A$ was determined. By doing this we were able to fit two sets of 14 data
points with two 3 parameter expressions. Overall the results are very
good for the lower and middle part of the 50-82 neutron shell. Even
though a particle-hole symmetry can be seen from the $\log(G)$ inset,
there is a discrepancy between the two as one moves towards the upper
part of the shell. This discrepancy is due to the shell closure, the
single-particle level structure, and the Pauli blocking for odd A nuclei.

\begin{figure}[htbp]
\includegraphics[width=8cm]{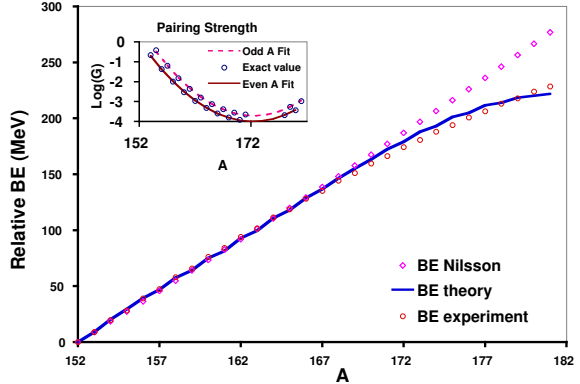}
\caption{The solid line gives the binding energies of Yb
isotopes relative to that of the $^{152}$Yb core. The single-particle
energy scale is set from the binding energy of $^{153}$Yb. The inset
shows the fit to values of $G$ that reproduce exactly the experimental
data. The two fitting functions are:
$\log(G(A))=662.2247 - 7.7912 A + 0.0226 A^2$ for even values of $A$
and
$\log(G(A))=716.3279 - 8.4049 A + 0.0244 A^2$ for odd values of $A$.
The Nilsson BE energy is the lowest energy of the non-interacting system.}
\label{Yb-isotopes}
\end{figure}

Figure \ref{Yb-isotopes} shows our results for the $^{154-161}$Yb 
isotopes.  We test the predictive power of the model on the 
$^{172-177}$Yb isotopes. In this case calculations of the pairing 
strength G are carried out only for the $^{154-171,178-181}$Yb 
isotopes and not for the $^{172-177}$Yb isotopes that are in the 
middle of the model space and thus are more computationally involved. 
This is a fit of two sets of 11 data points with two 3 parameter 
expressions. Then from the obtained quadratic polynomial fit to the 
$\log(G)$ values we calculate the theoretical values of the binding 
energy for these nuclei as shown in Figure \ref{Yb-isotopes}. This 
prediction is very good
when compared to the experiment. Thus, based on experimental data of the
nuclei in the upper and lower part of the shell and a $\log(G)$ fit 
to this data we can make a reasonable estimate for the mid-shell 
nuclei. Therefore, the model has a good predictive power.

\begin{figure}[htbp]
\includegraphics[width=8cm]{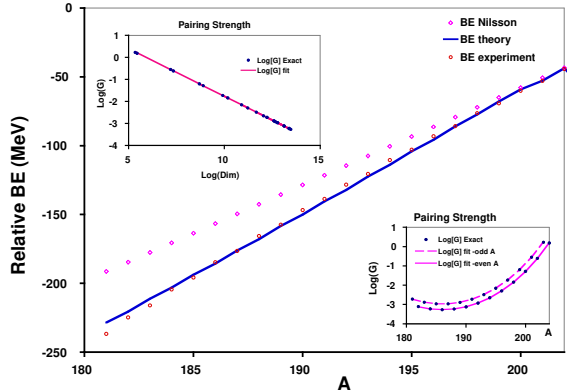}
\caption{The solid line gives the binding energy for
the Pb isotopes relative to the
$^{208}$Pb nucleus. The insets show the fit to the values of $G$ that
reproduce exactly the experimental data using a $^{164}$Pb core. The lower
inset shows the two fitting functions:
$\log(G(A))=382.3502 - 4.1375 A + 0.0111 A^2$ for even values of $A$
and
$\log(G(A))=391.6113 - 4.2374 A + 0.0114 A^2$ for odd values of $A$.
The upper inset shows a fit to $G(A)$ that is inversely proportional to
the size of the model space, ($\dim(A)$), that is valid for even as well
as odd values of $A$: $G(A)=366.7702/\dim(A)^{0.9972}$. The
Nilsson BE energy is the lowest energy of the non-interacting system.}
\label{Pb-isotopes}
\end{figure}

The next Figure \ref{Pb-isotopes} shows results for the $^{181-202 }$Pb
isotopes that were studied in the same way as the Sn and Yb isotopes. In
the calculations for these Pb isotopes, however, the binding energy that
was used is relative to that for $^{208}$Pb which was set to zero, but
the core nucleus was chosen to be $^{164}$Pb. Note that for the Yb and Sn
isotopes the core nucleus was also the binding energy reference nucleus
($^{100}$Sn and $^{152}$Yb).  In contrast, the calculation for the
Pb-isotopes was different because the core nucleus ($^{164}$Pb) and the
binding energy reference nucleus ($^{208}$Pb) are different. We again
have good quadratic fit to $\log(G)$ as function of $A$.

\begin{figure}[htbp]
\includegraphics[width=8cm]{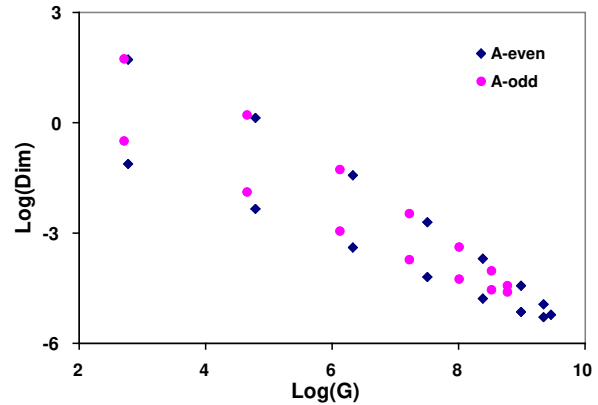}
\caption{Results for $\log(G)$
versa $\log(\dim)$ for the Sn nuclei.}
\label{Sn-log-log}
\end{figure}

The fact that there is a correlation between the pairing strength $G$
and the size of the model space reflected in the minimum of $G$ that is
at the maximal model space dimension prompted us to study $G(A)$ as
function of the model space dimension $\dim(A)$. In this respect, a
remarkable results is shown in Figure \ref{Pb-isotopes}. In this case the
pairing strength $G(A)$ for all the 21 nuclei (A=$181-202$) was fit by
a two parameter function $G(A)=\alpha /[\dim(A)]^{\beta}$ with the
values of the parameters taken to be $\alpha=366.7702$ and
$\beta=0.9972$. This function is inversely proportional to the
dimensionality of the model space $\dim(A)$. Unfortunately, this is not
the case for the other two isotopic chains. For example, in Figure
\ref{Sn-log-log} a $\log-\log$ plot is shown for the Sn nuclei.

It is unlikely that the linear relation between $\log(G)$ and
$\log(\dim(A))$ for the Pb isotopes is due to the difference between the
core and binding energy reference nucleus. What is more likely is that
this is due to the fact that in these cases the model space dimensions are
sufficiently large to result in a limiting form for the effective pairing
strength. Another reason might be that the single-particle energies and
the Pauli blocking mechanism are such that the even and odd parabolas of
$\log(G(A))$ produce a linear $\log-\log$ structure.

In the light of the above, it seems that the next step should be a study
that tracks the results as a function of the increasing size of the model
space to confirm or refute the $\log-\log$ relation. Such a study could also
address other questions such as the effect of the core binding energy as
a function of the deformation that is used in the Nilsson model to derive
the single-particle energies. Using a Woods-Saxon potential or other
methods to generate more realistic single-particle energies is another
opportunity for further studies.

In conclusion, we have studied binding energies of nuclei in three
isotopic chains: $^{100-130}$Sn, $^{152-181}$Yb, and $^{181-202 }$Pb
within the recently proposed extended pairing model \cite{FirstPaper} by
using Nilsson single-particle energies as the input mean-field energies.
Overall, the results suggest that the model is applicable to well-deformed
nuclei if the pairing strength is allowed to change as a (smooth) function
of the nucleon number $A$. The remarkably similar behavior of $\log(G)$
for even and odd $A$ values seems to suggest that there may be a single
$\log(G)$ function that bifurcates into an even-$A$ and an odd-$A$ branch
when the fermion dynamics is restricted to hard boson pairs only and Pauli
blocking is applied to exclude levels populated by the unpaired fermion.

\vskip .5cm Support provided by the U.S. National Science Foundation
(0140300), the Natural Science Foundation of China (10175031), and the
Education Department of Liaoning Province (202122024) is acknowledged.

\end{document}